\documentclass[12pt]{article}

\newcommand{\be}{\begin{equation}}
\newcommand{\ee}{\end{equation}}
\newcommand{\bea}{\begin{eqnarray}}
\newcommand{\eea}{\end{eqnarray}}
\newcommand{\nn}{\nonumber}
\newcommand{\w}{\omega}
\newcommand{\h}{\hbar}
\newcommand{\uw}{\underline{\omega}}
\newcommand{\ue}{\underline{\eta}}
\newcommand{\uc}{\underline{c}}
\newcommand{\uT}{\underline{\Theta}(\xi)}
\newcommand{\vv}{\mathrm{v}}
\newcommand{\ww}{\mathrm{w}}
\newcommand{\ahe}{$\bar{p}$He}

\begin{document}

\begin{titlepage}

\title{Semiclassical approach to the line shape}

\author{M.N. Stoilov\\
{\small\it Bulgarian Academy of Sciences,}\\
{\small\it Institute of Nuclear Research and Nuclear Energy,}\\
{\small\it Blvd. Tzarigradsko Chausse\'e 72, Sofia 1784, Bulgaria}\\
{\small e-mail: mstoilov@inrne.bas.bg}}

\maketitle

\begin{abstract}
We extend the results of Ref.~\cite{bjkst} on one-photon electric
dipole transition line shift and broadening to the case of
two-photon transitions. As an example we consider the laser
induced transition in antiprotonic helium produced in helium gas
target. The transition is between antiprotonic helium states
$(n,l)=(33,32)$ and $(31,30)$. \\ \\
PACS 32.70.Jz, 34.20.Gj, 36.10.-k
\end{abstract}
\end{titlepage}


\section{Introduction}
In the present paper we propose an approach to the evaluation of the density shift and broadening of the two-photon transition line profile.
The foundation of our considerations is the  method developed  in Ref.~\cite{bjkst} for $E1$-transitions.
We slightly modify the mathematical manipulations used therein and derive
close expressions for the shift of the resonance frequency and line broadening in terms of the perturbing potential mean value.

Our goal is to reduce the experimental uncertainty of recent
high-precision spectroscopy measurements \cite{hori} of
antiprotonic helium (\ahe), related to the effects of the
collisions of antiprotons with helium atoms. The example which we
consider thoroughly is  the two-photon transition from the initial
\ahe~ state $(n,l)=(33,32)$ to the final state $(31,30)$ \cite{asa}.
The transition is induced by external monochromatic
electromagnetic waves and is influenced by the helium gas target.
Unfortunately, the available data about the \ahe~ -- He interaction potential \cite{bjkst} do not cover the whole range of interparticle distances which we are interested in.
This forces us to use an extrapolation of the known potentials.
Two possible extrapolations are considered and compared.

\section{Two-photon transition in low density gas }

The general footing of our consideration is as follows: A quantum
system (emitter) is subject to a perturbation due to its randomly
propagating neighbors (perturbers) and also interacts with an
external electromagnetic field.
 The time dependent  Hamiltonian of the entire system is
\be
H(t)= H_0+V(t)+W(t)
\ee
where $H_0$ is the unperturbed Hamiltonian of the emitter,
$V$ is the emitter--perturbers potential, and $W$ is the electromagnetic interaction.

We suppose that
the emitter possesses  a full system of discrete states $\{ \vert j >\}$
with energies $\{ E_j\}$, i.e. $H_0 \vert j > = E_j \vert j >$.
In what follows
we shall distinguish three of these states, namely the successive states
$\vert i >, \vert m >$ and $\vert f >$ such that $E_i > E_m > E_f$.
The state $\vert i >$ will be our initial state and $\vert f >$ will be the final one.

In our consideration the electromagnetic field is a superposition  of
two plane waves with  frequencies $\w_k, \;\; k=1,2$.
The emitter --  electromagnetic field interaction   $W$ is of the form
\be
W(t)=  \ww^k cos( \w_k t).
\ee
The coefficients $\ww^k$ are emitter depending.
For example, if the emitter has a dipole moment $d$ then $\ww^k=d.E^k$
where $E^k$ are the electric field amplitudes.
It is supposed that the frequencies $\w_k$ are different but close to the  resonance ones
\bea
\w_1 &\neq& \w_2 \nn\\
\w_1 &\approx&  \w_{mi} = (E_i-E_m)/\h \nn\\
\w_2 &\approx&  \w_{fm} =(E_m-E_f)/\h .\label{ap1}
\eea

We make four assumptions for the potential $V$ \cite{bjkst}.
We suppose that, first, $V$ it is too weak to cause quantum
excitation in both emitter and perturbers. Second, the target
density is low enough, so that the emitter interacts only with one
perturber  at a time via the pairwise emitter -- perturber
interaction $V^0$. Third,  $V^0$  depends only on the distance $R$
between the emitter and perturber. Forth, we adopt  the Anderson
approach \cite{a1} and treat the perturbers classically. Moreover,
following \cite{bjkst}, we assume that the perturber's trajectory
$R(t)$ is entirely determined by the potential $V^0$ 
(plus initial conditions).

There are two important consequences of the above assumptions.
First, the perturbing potential $V$ is a sum of pairwise interactions
\be
V(t)=\sum_n V^0(R_n(t))
\ee
where $n$ runs over all perturbers.
Second, both $V^0$ and $V$ have the set $\{\vert j>\}$ as an eigenvector system
(with  $\vv^0_j(R)$ and $\vv_j(t)=\sum_n \vv^0_j(R_n(t))$ as eigenvalues).
Therefore, these operators commute with the Hamiltonian $H_0$
and the evolution operator of the initial system plus perturbation is diagonal with respect to the system of states $\{ \vert j>\}$.
Its matrix elements are determined by the action over the corresponding classical solution:
\be
<k\vert U_{H_0+V}(t_f,t_i) \vert j> =\delta_{kj}e^{-\frac{i}{ \h} E_j(t_f-t_i)
-i \eta_j(t_f,t_i)}.
\ee
Here
\be
\eta_j(t_f,t_i))=\frac{1}{ \h}\int_{t_i}^{t_f} d\tau \;\vv_j(t) \label{ac}
\ee
is the action of the perturbing potential when the emitter is in state
$\vert j>$.

The matrix element of the entire system evolution operator $U(t_f,t_i)$ between  states
 $\vert i>$ and $<f\vert$ up to the second order is
\bea
<f\vert U(t_f,t_i) \vert i> &=&
e^{-i(t_f-t_i)E_f/\h + i \eta_f(t_f,t_i)} [ \delta_{fi}\nn\\
&&
+\frac{ i}{ \h}  e^{i \w_{fi} t_i} \int_{t_i}^{t_f} d \tau\;
e^{-i \w_{fi}\tau -i\eta_{fi}(\tau,t_i)} W_{fi}(\tau) \nn\\
&&
-\frac{1}{\h^2} e^{i \w_{fi} t_i} \int_{t_i}^{t_f} d \tau_1\;
\sum_j e^{-i \w_{fj}\tau_1 -i\eta_{fj}(\tau_1,t_i)} W_{fj}(\tau_1)\nn\\
&&\times
\int_{t_i}^{\tau_1} d \tau_2\;
e^{-i \w_{ji}\tau_2 -i\eta_{ji}(\tau_2,t_i)} W_{ji}(\tau_2)].
 \label{u1}
\eea
Here $\eta_{kl}=\eta_l-\eta_k$ and $W_{kl}$ is the matrix element of the electromagnetic interaction operator $W$ between states
 $<k\vert$ and $\vert l>$.
We can simplify eq.(\ref{u1}).
We can neglect the common phase factor.
When $i \neq f$ we can omit the  diagonal term as well.
Keeping on only slowly oscillating terms we can reduce the sum over the full system of intermediate states to one term only, namely $\vert m>< m\vert$.
Finally, we can neglect by the same reason the first order term.
As a result the evolution operator matrix element between different states
 $<f\vert$ and $\vert i>$ takes the form:
\bea
<f\vert U(t_f,t_i) \vert i> &=&
\frac{\ww^2_{fm} \ww^1_{mi}}{\h^2}  \int_{t_i}^{t_f} d \tau_1\;
e^{-i (\w_{fm}-\w_2)\tau_1 -i\eta_{fm}(\tau_1,t_i)}\nn\\
&&\times
 \int_{t_i}^{\tau_1} d \tau_2\;
e^{-i (\w_{mi}-\w_1)\tau_2 -i\eta_{mi}(\tau_2,t_i)} .
 \label{u2}
\eea

It is clear that the line shape, shift and broadening are determined  by the double integral in eq.(\ref{u2}) and do not depend on the factor
 $\ww^2_{fm} \ww^1_{mi}/\h^2$.
In what follows we shall neglect this factor and concentrate our attention on the integral which in the limit
 $t_i \rightarrow -\infty \;\;\; t_f \rightarrow \infty $
we denote by $U$
\bea
U&=&\int d t\; e^{-i (\w_{fm}-\w_2) t -i\eta_{fm}(t)} I(t) \label{di}\\
I(t)&=& \int^t d\tau\; e^{-i (\w_{mi}-\w_1)\tau -i\eta_{mi}(\tau)}. \label{ii}
\eea

A few remarks. What we really know is $\vv^0_l(R)$ --- the
potential between the perturber and emitter in state $\vert l>$.
With proper boundary conditions (e.g., impact parameter\footnote{the distance between the emitter and the straight line which coincides with the perturber's trajectory at
infinity} $r$ and  velocity $v$) we can, using $\vv^0_i(R)$,
calculate the  perturber's trajectory $R(t)$. Knowing the
trajectory we can find $\vv^0_l(t;r,v)$ but this is not enough.
What we actually need is $\vv_l(t)$ which is a result of
successive random pairwise interactions. To find it we have to
choose a sequence of interaction moments $\{ t_{i}\}$ so that
$V(t)=\sum_i V^0(t-t_i;r_i,v_i)$. Using $V(t)$ we have to
calculate the probability of the process of interest and then to
average over all possible time sequences, impact parameters and
initial velocities. The task seems hopelessly complicated but
before to make some simplifications let us see what we can get.
 We expect, on the base of our assumptions about $V(t)$, that both $\eta_{fm}(t)$ and $\eta_{mi}(t)$ are stair-like functions of time, i.e.
(suppressing for a while the state indexes of $\eta$)
\be
\eta(t) = \eta_0 + c_0 t + \tilde\eta(t).\label{tac}
\ee
The important term here is $c_0 t$ while $\tilde\eta(t)$ gives a small irregular variation around it.
(The constant $\eta_0$ is irrelevant.
It gives an overall phase in the transition amplitude.)
Let us introduce the following simplifying notations:
\bea
\uw &=& \w_{fm}-\w_2+(c_0)_m-(c_0)_f\\
\w  &=& \w_{mi}-\w_1+ (c_0)_i-(c_0)_m\\
\ue &=& \eta_m -\eta_f\\
\eta&=& \eta_i -\eta_m
\eea
where $(c_0)_j$ is the $c_0$ coefficients of $\eta_j$.
We will need also the Fourier transform of $\tilde\eta$ and $\tilde{\ue}$
\be
\tilde\eta(t) = \int d \xi \;e^{i\xi t} \Theta(\xi),\;\;\;\;
\tilde{\ue}(t) = \int d \xi \;e^{i\xi t} \uT.
\ee
After regrouping the leading linear in $t$ terms in $I(t)$, the integral takes the form
\be
I(t)=\int^t d\tau e^{-i \w \tau -i\tilde\eta(\tau)}.
\ee
Now, because $V$ is small we suppose that $\tilde\eta(t)$ and $\tilde{\ue}$ are small as well (see the comment below about this point), so we can make series expansion of
$e^{-i\tilde\eta}$ and $e^{-i\tilde{\ue}}$.
Therefore, $I(t)$ and $U$ can be rewritten as follows
\bea
I(t)&=&
\int^t d\tau \;e^{-i \w \tau}\left(1 -i\tilde\eta(\tau)\right)\nonumber\\
&=&e^{-i \w t}
\left(\frac{i}{\w} + \int d\xi\;\frac{\Theta(\xi)}{\w-\xi}e^{i\xi t}\right)\label{ise}
\eea
\bea
U&=&\int d t\; e^{-i ( \uw + \w) t}\left(1-i\tilde{\ue}(t)\right)
\left(\frac{i}{\w} + \int d \xi \; \frac{\Theta(\xi)}{\w-\xi}e^{i\xi t}\right)\nn\\
&=&\int d t\; e^{-i ( \uw + \w) t}
\left(\frac{i}{\w} +
\int d \xi\left(\frac{\uT}{\w}+
 \frac{\Theta(\xi)}{\w-\xi}\right) e^{i\xi t}\right)\nn\\
&=&2\pi\left(\frac{i \delta(\uw+\w)}{\w} +\frac{\underline{\Theta}(\uw+\w)}{\w}
-\frac{\Theta(\uw+\w)}{\uw}\right).\label{u3}
\eea
The line shift $\Delta$ of the two-photon transition can be read immediately   from eq.(\ref{u3}).
Recalling the definitions of $\uw$ and $\w$ we see that
$
\uw+\w= (E_i-E_f)/\h -\w^1 -\w^2 +  (c_0)_i - (c_0)_f
$
and therefore
\be
\Delta = (c_0)_i - (c_0)_f. \label{shif1}
\ee
Note that
\be (c_0)_l=\lim_{t\rightarrow\infty} \frac{1}{2 t}\int_{-t}^t d\tau\;\vv_l(\tau),
\label{mvpp}
\ee
i.e., $c_0$ is the mean value of the perturbing potential.

It is easy to find the line shift, but the situation with the line broadening is more complicated.
At the present moment we can not say anything about it, because we do not know the explicit form of $\Theta(\xi)$ and $\uT$.
We shall devote the rest of the paper to show that for small but non-zero $\xi$ the functional form of $\Theta$ (and $\underline{\Theta}$) is entirely determined by the coefficient $c_0$ and it is
\be
\Theta(\xi)\propto\frac{c_0}{\xi}\label{asymp}.
\ee

We want to step back a little and comment the expansion of 
$e^{-i\tilde\eta(t)}$ used in eqs.(\ref{ise},\ref{u3}).
The assumption that $\tilde\eta$ is small for any $t$ is  correct, if we have a gap (or cut off) near the zero in the spectrum of $\tilde\eta$.
The easiest may to ensure the existence of a gap is to suppose that $V(t)$ is periodic.
The idea  is that the  mean value of the perturbing potential which determines both the shift and broadening of the line
is independent,  according to the Central Limit Theorem, of the assumption for periodicity, but can be easily calculated using it.

Supposing that the perturbation is periodic its period  $t_0$ is the time between two emitter -- perturber impacts. 
This time can be determined by the density $N$ and temperature $T$ of the perturbers
\be
t_0= \frac{1}{N \;\pi  r_{max}^2\; \bar v}\label{tim1}
\ee
where $\bar v = \sqrt{8 k T/\pi m}$ is the mean speed of the perturbers ($m$ is the reduced mass of the system) and $r_{max}$ is the distance above which we can neglect the perturbation.
Obviously, $r_{max}$ depends on $V^0$ but is also in our hands.

The assumption of periodicity  leads to a dramatic simplification in the calculations.
In the periodic picture
\be
V(t)=\sum_n \bar{V}^0(t+n t_0)
\ee
where $\bar V_0(t)$ is the mean emitter -- perturber potential
\bea
\bar V^0(t) &=& \frac{1}{r_{max}^2}\int^{r_{max}} d r \;2 r\;
\sqrt{\frac{2}{\pi}\left(\frac{m}{k T}\right)^3}\nn\\
&&\;\;\;\times
\int d v\; e^{-m\;v^2/2k\; T} v^2 V^0(t;r,v). \label{avp}
\eea
Certainly, $\bar V_0(t)$ commutes with the Hamiltonian.
We denote its eigenvalues with $\bar\vv^0_l$.
The time parameter in eq.(\ref{avp}) is chosen so that $t=0$ corresponds to the apex of the perturber's trajectory.
Therefore,  $\bar\vv^0_l(t)$ and $\vv_l(t)$ are symmetric functions of $t$.
For $\vv_l(t)$ we can write down a Fourier series
\bea
\vv_l(t)/\h  &=& (c_0)_l + 2 (c_k)_l \cos(k \w_0 t).\label{pfc}\\
(c_k)_l      &=& \frac{1}{t_0\h}\int_{-t_0/2}^{t_0/2}d t\;
                    \cos(n \w_0 t) \bar \vv^0_l(t).
\eea
Note that, because of the existence of $r_{max}$, what we really calculate is
\be
(c_k)_l = \frac{1}{t_0\h}\int_{-t_1/2}^{t_1/2}d t\; \cos(k \w_0 t) \bar \vv^0_l(t)
\label{fc}
\ee
where $t_1$ is the time for which the perturber propagates through the area of nonzero potential.
According to one of our assumptions listed above $t_0 >> t_1$.

Eq.(\ref{fc}) is an origin of a very useful symmetry
\bea
t_0 &\rightarrow &\alpha t_0\nn\\
\bar V^0(t) &\rightarrow &\alpha \bar V^0(t)\label{scal}
\eea
provided $\alpha t_0 > t_1$.
Proof:
Let us change $t_0$ so that $t_0^{new}=t_0/k$.
Therefore, $\w_0^{new}=k \w_0$ and
\be
c_n^{new} = \frac{k}{t_0\h}\int_{-t_1/2}^{t_1/2} \cos(n k\w_0 t) \bar\vv^0(t).
d t = k c_{k n}
\ee
Coefficients $c_n^{new}$ and $c_{kn}$ correspond to one and the same frequency
$n \w_0^{new}= k n \w_0$ which means that
$
c^{new}(\w) = k c(\w).
$
If we rescale both $t_0$ and $\bar V^0$ as it is prescribed by eqs.(\ref{scal}) then we will get that $c^{new}(\w)=c(\w)$.
Now look at eqs.(\ref{tim1}, \ref{avp}).
We see that different choice of $r_{max}$ leads exactly to the transformation (\ref{scal}).
Increasing $r_{max}$ we get weaker mean perturbation but it happens more often with the same gross effect.
When $1/\alpha$ is not integer there will be a shift in the overtone positions, but still the Fourier coefficients will lay on one and the same curve.
This curve - the envelope of the Fourier coefficients is the important one for us and it determines the line shape.
Note that symmetry (\ref{scal}) holds both for $\alpha <1$ and $\alpha > 1$.
In the latter case the only limit on $\alpha$ is determine by the condition
 $\bar {V^0}^{new}/\h <<1$ no matter to what $r_{max}$ it corresponds.
We shall use such transformation to probe the line shape at small frequencies.

Using eq.(\ref{pfc}) we get  the following expressions for the actions $\eta$ and $\ue$ (see also eq.(\ref{tac})):
\bea
\eta(t) &=& c_0 t + \sum_{k=1} \frac{2 c_k}{k \w_0} \sin(k \w_0 t))\\
\ue(t) &=& \uc_0 t + \sum_{k=1} \frac{2 \uc_k}{k \w_0} \sin(k \w_0 t)).
\eea
 In the right hand side of the above equations all therms are small\footnote{this gives another way to determine the maximal $\alpha$ (or the minimal $\w_0$) we can use} but the first ones (because $t$ is arbitrary).
 Therefore
\be
e^{-i\eta(t)}= e^{-i c_0 t}(1 - i \frac{2 c_k}{k \w_0} \sin(k \w_0 t))
\ee
and
\bea
I(t) &=&
i e^{-i \w t} \left(\frac{1}{\w}
-\frac{2 c_k}{{\w}^2-(k\w_0)^2} \cos(k \w_0 t) \right.\nn\\
&&- \left. i \frac{2 c_k\w }{k \w_0({\w}^2-(k\w_0)^2)}\sin(k \w_0 t)\right)
\eea
Now we shall need some well known formulas
\bea
\int_{-\infty}^\infty d t\;e^{-i \w t} \cos(\bar\w t) &=&
\pi\left( \delta(\w+\bar\w)+\delta(\w-\bar\w)\right)\\
\int_{-\infty}^\infty d t\;e^{-i \w t} \sin(\bar\w t) &=&
i \pi \left(\delta(\w+\bar\w)- \delta(\w-\bar\w)\right)
\eea
Using these equations we obtain that $U$ (up to a phase) is
\bea
U
 &=& 2\pi  \left( \frac{1}{\w }\delta(\uw+\w) +\right.\nn\\
&& +\left(\frac{c_k}{\uw (\uw+\w) } -
 \frac{\uc_k}{\w (\uw+\w)}\right)\delta(\uw+\w - k\w_0) +\nn\\
&& +\left.\left(\frac{c_k}{\uw (\uw+\w)} -
 \frac{\uc_k}{\w (\uw+\w)}\right)\delta(\uw+\w + k\w_0) \right)
 \label{u4}
\eea
Hereafter we suppose that one of the frequencies $\w^k$, say $\w^1$, is fixed.
Then the probability interpretation of eq.(\ref{u4}) is  exactly the same as for the simple one-photon transition amplitude between unperturbed states.
The only difference is that  instead of one line now
we have a bunch of closely separated lines with intensities  proportional to the square of the coefficients in front the delta functions.
In practice, where the real potential is not periodic, we see the envelope of these lines.
This is in agreement with eqs.(\ref{fc},\ref{scal}) according to which at the limit
$t_0\rightarrow\infty$ the inter line distance $\w_0$ is so small that the Fourier coefficients $c_k$ form a line.
The important moment is that the Fourier coefficients calculated for any $t_0$ lay on this line.
Therefore, we can fix $t_0$, find the Fourier coefficients $\{c_k\}$ and $\{\uc_k\}$  and interpolate them with functions $c(\xi)$ and $\uc(\xi)$.

The expression for $U$ given in eq.(\ref{u4}) is very close to that in eq.(\ref{u3}).
It is clear that the line shift again is determined by eq.(\ref{shif1}) but now we can say something more about line shape.
The envelope functions $c(\xi)$ and $\uc(\xi)$, according to
  eq.(\ref{fc}),  behave like constants  for $\xi \rightarrow 0$.
So, we can smoothly continue them for negative $\xi$ making them symmetric with respect to $\xi=0$.
As a result $U$  for sufficiently small but non zero $\xi$ is  (see also eq.(\ref{asymp}))
\be
U\propto
\frac{1 }{(\uw+\w)}\left(\frac{c_0}{\uw}-\frac{\uc_0}{\w}\right)
\label{frm}
\ee
Eq.(\ref{frm}) allows us to estimate for given $\w^1$  the line broadening $\sigma$ (calculated as the difference between frequencies for which the probability is half of its maximum)
\be
\sigma=\sqrt{(\w \pm \sqrt{2} \uc_0)^2 \pm 4 \sqrt{2} c_0 \w}.\label{brdp}
\ee
The choice of the sign in eq.(\ref{brdp}) depends on which of discriminants
is positive.
If both  discriminants are positive then we have double line.
 (The doublet  can be seen easily as one much broader line in experiments with non monochromatic light sources.)

The  dependence of the Fourier coefficients $c$ on density $N$ can be read from eqs.(\ref{tim1},\ref{fc})
\be
c_k(N) = \left(\frac{c_k(N_0)}{N_0}\right) N.\label{ddc}
\ee
As a consequence of eq.(\ref{ddc}) the line shift and broadening also depend on $N$
\bea
\Delta(N) &=& \left( \frac{c_0(N_0)+\uc_0(N_0)}{N_0}\right) N \label{ls}\\
\sigma(N) &=& \left( \frac{\sqrt{2}\vert\uc_0(N_0)\vert}{N_0}\right) N 
\;\;\; \mathrm{if}\;\;\vert\uc_0\vert,\;\vert c_0\vert >> \vert\w\vert
\label{br1}\\
\sigma(N) &=& \vert\w\vert \hspace{29 mm} \mathrm{if}\;\;\vert\uc_0\vert,\;\vert c_0\vert << \vert\w\vert 
\label{br2}.
\eea

\section{Numerical results}

The system we consider \cite{asa} consists of  antiprotonic helium as emitter and the helium atoms  in a gas target as perturbers.
The initial, final and intermediate states are
$\vert i>= (33,32)$, $\vert f>= (31,30)$ and  $\vert m>= (32,31)$ respectively.
The target is  at $p=1$ mbar and $T=6^{\mathrm o}$K.

We use two sets of perturbing potentials $\{\vv^0_i, \vv^0_m$ and $\vv^0_f\}$
corresponding to two different extrapolations of the data we have about the potential energy surface (PES) for the \ahe~ -- He interaction.
The results obtained from the first set of potentials will be
indicated by prime and those obtained from the second set by
double prime.
%

The double integral in Eq.~(\ref{avp}) was calculated as a left Riemann
sum over a regular set of $N=1000$ points for the impact parameter
$r\le r_{max}=25$ a.u. and using a Gauss-type quadrature formula
with $M=6$ points for the average over the Maxwell distribution
for the velocity. %

We obtain the following values for 
the coefficients $c_0$ and $\uc_0$ needed to estimate the line shift and broadening according to eqs.(\ref{ls} -- \ref{br2}) 
\bea
c_0'(N_0)/N_0   &=& 5.2 \;\;10^{-13} [\mathrm{Hz\;cm^{3}}]\\
\uc_0'(N_0)/N_0 &=& 4.8 \;\;10^{-13} [\mathrm{Hz\;cm^{3}}]\\
&&\nonumber\\
c_0''(N_0)/N_0    &=& 1.4 \;\;10^{-12} [\mathrm{Hz\;cm^{3}}]\\
\uc_0''(N_0)/N_0  &=& 1.4 \;\;10^{-12} [\mathrm{Hz\;cm^{3}}].
\eea
This gives that the line shift is
\bea
\Delta'(N_0)/N_0  &=& 1.0 \;\;10^{-12} [\mathrm{Hz\;cm^{3}}]\\
\Delta''(N_0)/N_0  &=& 2.8 \;\;10^{-12} [\mathrm{Hz\;cm^{3}}].
\eea
We can use eq. (\ref{br1}) to obtain the line broadening when 
$ \vert \w_1\vert \approx \vert \w_{mi}+c_0\vert $, 
i.e., when we have a fine tuning between the first laser frequency and the inter level distance between the initial and intermediate states. 
Then the result is
\bea 
\sigma'(N_0)/N_0  &=& 6.8 \;\;10^{-13} [\mathrm{Hz\;cm^{3}}]\\
\sigma''(N_0)/N_0 &=& 2.0 \;\;10^{-12} [\mathrm{Hz\;cm^{3}}].
\eea
Eq.(\ref{br2}) describes the off-resonance situation.
In this case the line broadening  does not depend on the target density  
\be 
\sigma(N) = \vert \w_{mi}-\w_1\vert \;\;\;\; \mathrm{if}\;\;\vert \w_{mi}-\w_1\vert >> \vert\uc_0\vert,\;\vert c_0\vert.
\ee
In every other case the general formula (\ref{brdp}) for the line broadening has to be used.

 The discrepancy between results for the two approximating
sets of potentials is about a factor of three. This indicates that
the PES extrapolation we have used in the
construction of the potentials is not reliable. The problem could
be solved only by extending the PES to shorter distances between
the antiproton and He$^+$ ion, corresponding to the average
radius of the antiproton orbit in \ahe~ in states with
$n\sim30$.

\section*{Acknowledgments}
The author is grateful to Dr. D.Bakalov for the helpful
discussions and to the Organizing committee of the PSAS'2010 Workshop.
 The work was supported in part by BNSF under grant 2-288.

\end{document}